\documentclass{PoS}
\usepackage{graphicx}
\def\gb{\texttt{ghost busters}}

\title{In-flight calibration of the INTEGRAL/IBIS mask}

\ShortTitle{In-flight calibration of the INTEGRAL/IBIS mask}

\author{\speaker{Simona Soldi}$^{a,b}$, Fran\c{c}ois Lebrun$^b$, Aleksandra Gros$^a$, Guillaume B\'elanger$^c$, Volker Beckmann$^{b,d}$, 
	Isabel Caballero$^a$, Andrea Goldwurm$^{a}$, Diego G\"otz$^a$, Fabio Mattana$^{b,d}$, Juan A. Zurita Heras$^{b,d}$, Angela Bazzano$^e$, Pietro Ubertini$^e$
         \\
        $^a$ Laboratoire AIM, CEA/IRFU, CNRS/INSU, Universit\'e Paris Diderot, CEA DSM/IRFU/SAp, 91191 Gif-sur-Yvette, France \\
	$^b$ APC, Universit\'e Paris Diderot, CNRS/IN2P3, CEA/Irfu, Observatoire de Paris, Sorbonne Paris Cit\'e, 10 rue Alice Domon et L\'eonie Duquet, 
	     75205 Paris Cedex 13, France \\
	$^c$ European Space Agency (ESA) / ESAC - P.O. Box 778, Villanueva de la Canada, 28691 Madrid, Spain \\
	$^d$ Fran\c{c}ois Arago Centre, APC, Universit\'e Paris Diderot, CNRS/IN2P3, CEA/Irfu, Observatoire de Paris, Sorbonne Paris Cit\'e, 13 Rue Watt, 75013 Paris, France \\
	$^e$ INAF-Istituto di Astrofisica e Planetologia Spaziali, Via del Fosso del Cavaliere 100, I-00133, Roma, Italy \\
        E-mail: \email{soldi@apc.univ-paris7.fr}}

\abstract{
Since the release of the \textit{INTEGRAL} Offline Scientific Analysis (OSA) software version 9.0, the \gb\ module has been introduced in the
\textit{INTEGRAL}/IBIS imaging procedure, leading to an improvement of the sensitivity around bright sources up to a factor of 7.
This module excludes in the deconvolution process the IBIS/ISGRI detector pixels corresponding to the projection of a bright source 
through mask elements affected by some defects. These defects are most likely associated with screws 
and glue fixing the IBIS mask to its support.
Following these major improvements introduced in OSA~9, a second order correction is still required to further remove the residual 
noise, now at a level of 0.2--1\% of the brightest source in the field of view. In order to improve our knowledge of the IBIS mask transparency, 
a calibration campaign has been carried out during 2010--2012. 
We present here the analysis of these data, together with archival observations of the Crab and Cyg~X--1, that allowed us to build a composite 
image of the mask defects and to investigate the origin of the residual noise in the IBIS/ISGRI images. 
Thanks to this study, we were able to point out a simple modification of the ISGRI analysis software that allows to significantly improve the quality of
the images in which bright sources are detected at the edge of the field of view. Moreover, a refinement of the area excluded by the \gb\ module
is considered, and preliminary results show improvements to be further tested. 
Finally, this study indicates further directions to be investigated for improving
the ISGRI sensitivity, such as taking into account the thickness of the screws in the mask model or studying the possible discrepancy between
the modeled and actual mask element bridges.
}

\FullConference{ An INTEGRAL view of the high-energy sky (the first 10 years) - 9th INTEGRAL Workshop and celebration of the 10th anniversary of the launch\\
                 15-19 October 2012\\
                 Bibliotheque Nationale de France, Paris, France}

\begin{document}

\section{Introduction}
\vspace{-0.3cm}
The IBIS instrument on board the \textit{INTEGRAL} satellite \cite{winkler03} has been designed to image the hard X-ray sky in the energy range between 15~keV to 10~MeV 
making use of the coded mask technique \cite{ubertini03}. The IBIS mask is made of 11.2$\times$11.2$\times$16 $\rm mm^3$ tungsten elements and it is
based on a truncated replication of a Modified Uniformly Redundant Array (MURA).
Since the effective area depends on the number of transparent mask elements seen by the detector and since in the gamma-ray band the count rate is dominated 
by the internal background, the best compromise is achieved with half of the elements being opaque and the other half transparent \cite{skinner08}.

\begin{figure}[!t]
\hspace{4cm}
\includegraphics[width=0.42\textwidth]{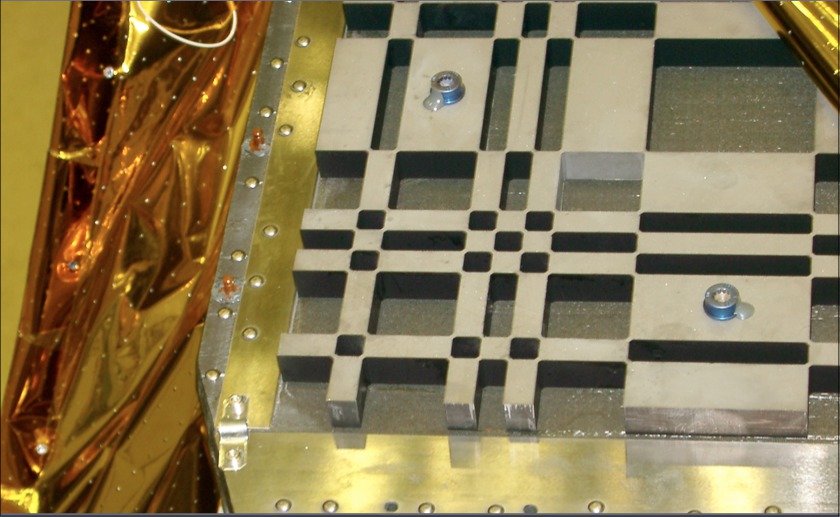} 
\caption{A section of the flight model of the IBIS mask during the integration phase, showing two of the screws used to fix the mask to its support and some overflowing glue
(Credits ESA).} 
\label{glue}
\end{figure}

The Offline Scientific Analysis (OSA) software package provides the complete set of tools to analyse the \textit{INTEGRAL} data.
Within the IBIS analysis software \cite{goldwurm03}, the coded pattern and the support structures of the IBIS mask (e.g. the honeycomb support structure known as ``nomex'') 
are modeled and therefore their contribution is taken into account during the data reduction. 
However, the imaging software currently available in OSA~10 (which is the same as in OSA~9) does not take into account the irregularities in the mask assembly 
and the variations in transparency caused, for example, by screws and glue that have been used to fix the mask to its support (Fig.~\ref{glue}). 
These components provide additional 
absorption of the source photons, therefore being at least partly responsible for residual coding noise (ghosts\footnote{For an on-axis point source, the ISGRI point spread function 
presents a central peak at the true position of the source in the image, plus 8 secondary, spurious peaks (ghosts) in well defined positions (determined 
by the size of the basic pattern), which are cleaned during the image reconstruction process \cite{gros03}.}) in the IBIS/ISGRI \cite{lebrun03} images and limiting the instrument 
sensitivity, especially when bright sources are present in field of view (FOV).

\begin{figure}[!b]
\includegraphics[width=1.02\textwidth]{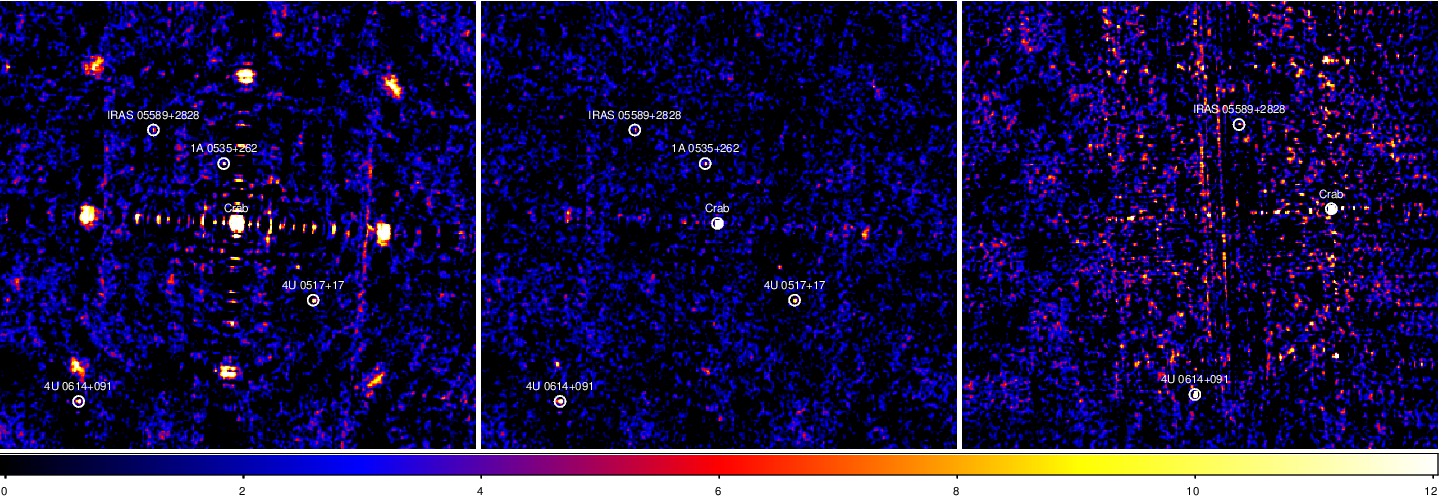} 
\caption{\footnotesize{20--60~keV ISGRI significance images of the region around the Crab. \textit{Left}: OSA~8, Crab on-axis. 
\textit{Centre}: OSA~9 (same data as in the left panel). \textit{Right}: OSA~10 for off-axis Crab observations. A major improvement is evident between OSA~8 and OSA~9/10, but residual noise, 
especially for data with bright off-axis sources, is still present. \texttt{[See http://pos.sissa.it/cgi-bin/reader/conf.cgi?confid=176 for the full resolution version of this figure]}}} 
\label{ghost}
\end{figure}

Since OSA~9, the \gb\ routine (developed by D. Eckert and collaborators at ISDC\footnote{\emph{http://www.isdc.unige.ch/newsletter.cgi?n=n22}}) 
automatically excludes from the deconvolution 
process the ISGRI detector pixels corresponding to the projection of a bright source through mask elements affected by these defects. 
This entails a major improvement in the ISGRI imaging capabilities (see Fig.~\ref{ghost}, left and central panels). Nevertheless, a second 
order correction is still required to remove further residual noise from the images, now at a level of 0.2--1.0\% of the brightest source in the FOV 
(see Fig.~\ref{ghost}, right panel).

\section{Imaging of the \textit{INTEGRAL} IBIS mask}
\vspace{-0.3cm}
To reconstruct a truly accurate model of the mask assembly, it is necessary to measure the variations of transparency caused by the unmodeled components
of the mask.
Our estimate indicates the need for about 1~Ms exposure on the edges of the mask in order to characterize its transparency to an accuracy of about 1\%.

\begin{figure} 
\hspace{1.5cm}
\includegraphics[width=0.79\textwidth]{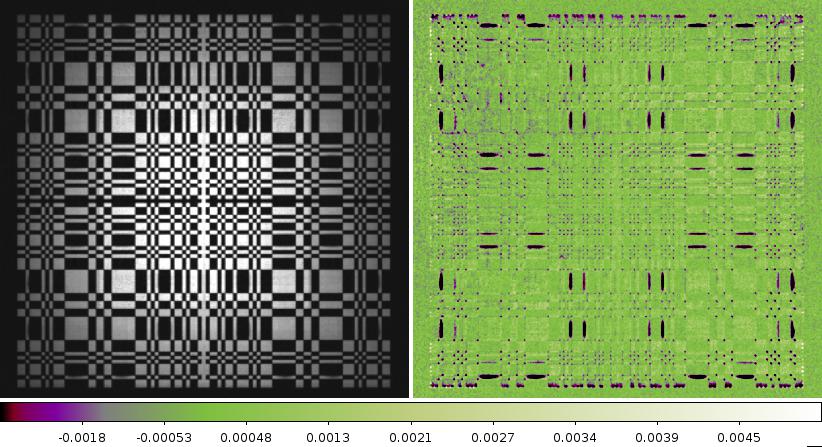} 
\caption{\footnotesize{\textit{Left}: Sum of the back-projected ISGRI shadowgrams for all the available Crab and Cyg X--1 data (up to revolution 1141; 18--60~keV range), 
reproducing a composite radiography of the IBIS mask, where all observing directions are combined. 
\textit{Right}: composite radiography of the mask defects. Negative values correspond to regions more absorbed in the real mask 
than in the OSA mask model. These zones are associated to the presence of screws and glue fixing the mask to its support. 
In addition, the dotted structure in the image corresponds to the position of some of the bridges between mask elements.
\texttt{[See http://pos.sissa.it/cgi-bin/reader/conf.cgi?confid=176 for the full resolution version of this figure]}}} 
\label{radio}
\end{figure}

The technique chosen to image the defects of the mask consists in using bright and relatively isolated sources, such as the Crab and Cyg X--1,
to reconstruct a composite image of the IBIS mask by back-projecting from the ISGRI detector to the mask plane the shadowgrams produced by these sources
(left panel in Fig.~\ref{radio}).  
Subtracting this composite image from the mask model highlights the unmodeled defects of the mask (right panel in Fig.~\ref{radio}).
We refer to this measurement as \textit{radiography} since we image the mask ``from behind'' using the light modulated by the mask and detected by ISGRI. 
When interpreting the radiography of the mask defects, it is important to take into account that all observing directions 
are combined to obtain a complete image of the mask. 
As a result of this, for example the screws closer to the mask borders appears with a rounded external edge (right panel in Fig.~\ref{radio}).

\subsection{Calibration observations}
To build the radiography of the mask defects, we collected all Crab and Cyg~X--1 pointings up to revolution 1141 (February 18, 2012), 
with an exposure time larger than 800~s and in which the source was within a $20^{\circ}$ radius from the pointing direction. 
These data include a dedicated calibration program designed to observe the less exposed corners of the mask, performed during 2010--2012 for a total 
of 5 revolutions. Each observation of a mask corner was executed with a custom pattern composed by 3 grids of 4$\times$4 pointings.
The characteristics of the pattern (2.33$^{\circ}$ separation between the grid centers, 20' step between the pointings) and the pointing direction were 
chosen in order to maximize the exposure of the mask borders 
and to obtain data that could still be usable by SPI for calibration purposes.
This program allowed us to accumulate $\sim700 \rm \, ks$ of effective exposure centered on the mask edges.

In addition, following the PI's approval, scientific observations of the Cygnus region during AO--7 have been adapted, in order to contribute to 
the mask calibration (i.e., the satellite Z-axis has been optimized to observe the less exposed corners of the mask, without affecting the science goals
of the program). Unfortunately, during these observations at the end of 2010 (and further until March 2011, and then again in autumn 2011)
the hard X-ray flux of Cyg~X--1 dropped to a level of about 20\% of its average value in the 2005--2010 period, therefore significantly limiting the contribution
of these observations to the mask radiography data set.

The radiography of the mask defects shown in Fig.~\ref{radio} and the results presented here are therefore based on all available data from the beginning
of the mission up to the end of 2011. Including also the additional dedicated observation performed in 2012 during about one revolution, three mask corners
result to have $\sim 700 \rm \, ks$ of effective exposure time, while the remaining 4th corner has $\sim 500 \rm \, ks$. 
This allows us to map the IBIS mask transparency with unprecedented accuracy.

\subsection{First results and future work}

The first results from this on-going effort show the need to increase the number of detector pixels excluded by the \gb\ routine, and the need to apply \gb\
also to sources located at the very edge of the ISGRI FOV.
In fact, when the radiography of the mask defects is built using shadowgrams on which \gb\ has been applied, part of the screws and of the structures 
at the mask borders can still be observed, pointing out that 1) the limits in FOV pixel coordinates for a source to be treated with \gb\ have to be increased
(from [--160,160] to [--180,180]), and 2) a larger area at the upper and lower mask borders, and around the screws located closest to the borders needs to be 
filtered out.

\begin{figure} 
\hspace{1.8cm}
\includegraphics[width=0.70\textwidth]{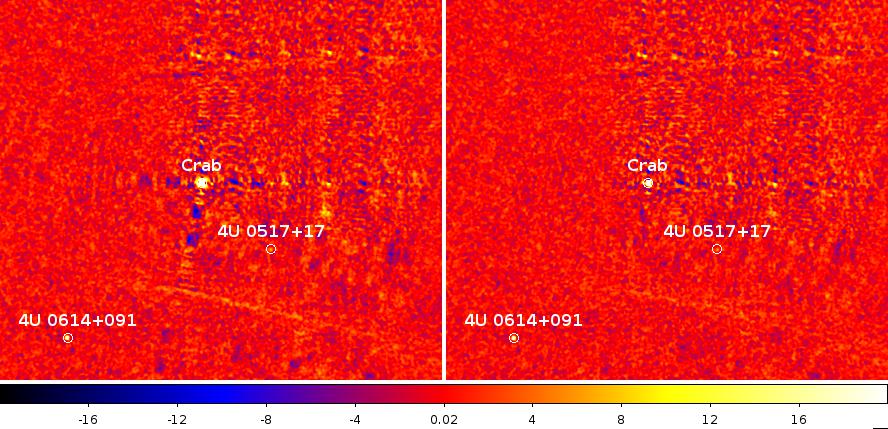} 
\caption{\footnotesize{20--60~keV ISGRI significance images of the region around the Crab (off-axis angles 13--20$^{\circ}$, 
images normalized to the peak significance of the Crab in OSA~10). Left: OSA~10. Right: OSA~10 with the modified \gb\ (OSA~10$+$), 
as explained in the text. \texttt{[See http://pos.sissa.it/cgi-bin/reader/conf.cgi?confid=176 for the full resolution version of this figure]}}} 
\label{offaxis}
\end{figure}

Indeed, when the \gb\ software is modified to correct for sources positioned all the way out to the edge of the ISGRI FOV, a significant improvement 
is observed in the ISGRI images. In Fig.~\ref{offaxis} we show the ISGRI images obtained with the standard OSA~10 software (left panel) for all pointings
where the Crab is at off-axis angles between 13--20$^{\circ}$ (112 pointings, out of which 67 do not have \gb\ applied within OSA~10). The right panel
shows the improvement when \gb\ is applied to all pointings (hereafter OSA~10$+$). When fitting the pixel significance distribution with a Gaussian function, the distribution 
width is found to decrease from $\sigma_{Gauss, OSA10} = 1.363 \pm 0.002$ to $\sigma_{Gauss, OSA10+} = 1.337 \pm 0.002$. The most prominent noise 
structures have decreased their significance by a factor of 2--5 and the total number of image pixels with absolute value larger than $10 \sigma$ decreased by 
45\%.
Similar results are obtained for Cyg~X--1, when using all pointings with off-axis angles between 13--14$^{\circ}$ (397 pointings, out of which 145 do not 
have \gb\ applied within OSA~10), with the distribution width of the pixel significances decreasing from 
$\sigma_{Gauss, OSA10} = 1.36277 \pm 0.00007$ to $\sigma_{Gauss, OSA10+} = 1.3417 \pm 0.0006$ and the total number of image pixels with absolute value larger 
than $10 \sigma$ decreased by 20\%.

Moreover, we made a preliminary study to test a different OSA modification, i.e., that of the mask regions that are filtered out by \gb\ (hereafter OSA~10$_{\rm test}$).
We selected a data set with the Crab at off-axis angles mostly between 7--16$^{\circ}$ for which significant noise structures are still present in 
the ISGRI image at a level of 0.2--1.0\% of the Crab (same data as in the right panel of Fig.~\ref{ghost}). 
While the distribution width of the pixel significance does not vary, some of the localized noise does decrease. Indeed, the total number of image pixels 
with absolute value larger than $10 \sigma$ decreased by 14\% when increasing the excluded mask area. Moreover, when subtracting the two significance maps 
(OSA~10 -- OSA~10$_{\rm test}$) and counting the pixels for which a variations larger than 2$\sigma$ was observed (excluding the pixels at the position
of detected sources), we see that 65\% of those decreased their significance in OSA~10$_{\rm test}$ against 35\% that increased it, pointing towards
a global net improvement in the ISGRI images.
Further tests on different data sets, and in crowded regions of the sky, will be performed before the next release of the OSA software.
The energy dependence of the mask defects has not been taken into account for these tests, and will be the object of a future study.

Finally, the radiography of the mask defects gives a hint that the documented and modeled bridges between opaque, adjacent mask elements 
(Fig.~\ref{glue}) might not correspond to the actual mask construction. This could be the reason for the bridge residuals visible in the 
mask defect radiography for observations with the source on-axis. On the other hand, the increase of these residuals with the off-axis angle requires 
a deeper analysis (see the size increase of the dots when going towards the mask corners in Fig~\ref{radio}, right panel).
In addition, we are currently testing a further improved version of \gb, which takes into account the thickness of the mask screws. 
This modification will allow us to apply a more detailed pixel exclusion, minimizing the rejected signal, which is especially important for crowded fields 
with several bright sources. 

\vspace{-0.3cm}
\section{Conclusions}
\vspace{-0.3cm}
A dedicated calibration campaign has been carried out by \textit{INTEGRAL} in 2010--2012 in order to map the transparency of the IBIS mask and of those 
supporting structures which are not currently included in the OSA software. 
Combining these with archival data and imaging the mask and its defects, we were able to identify a modification of the ISGRI imaging software --
in particular within the \gb\ routine -- that will result in a global decrease of the noise structures in the ISGRI images when bright sources are observed 
at the edge of the field of view. The modified routine will be included in the next release of the OSA software. \\
A further promising modification, concerning the detector area which is excluded from deconvolution due to the presence
of irregularities in the mask assembly, is currently under study, together with the possibility of a finer mask modelling, which takes into account the 
thickness of the mask screws and the energy dependence of the mask defects.

\vspace{-0.3cm}
\section*{Acknowledgments}
\vspace{-0.3cm}
\small{
We wish to thank Pierrick Martin (IPAG, Grenoble) for allowing the modification and use of his Cygnus observations by the calibration 
team, the INTEGRAL User Group for supporting the mask calibration, Erik Kuulkers and Celia S\'anchez-Fern\'andez (ESAC, Madrid) for helping
preparing the observation pattern and scheduling this dedicated program, the SPI and JEM-X teams for permitting this calibration campaign. 
SS, IC, FM and JAZH acknowledge the Centre National d'Etudes Spatiales (CNES) for financial support.
The present work is based on observations with \textit{INTEGRAL}, an ESA project with
instruments and science data centre funded by ESA member states (especially the PI countries: Denmark, France, Germany,
Italy, Switzerland, Spain, Czech Republic and Poland, and with the participation of Russia and the USA).
ISGRI has been realized and is maintained in flight by CEA-Saclay/Irfu with the support of CNES.\\
We thank the anonymous referee for the constructive comments that helped to improve this paper.
}

\end{document}